\documentclass[journal]{IEEEtran}
\IEEEoverridecommandlockouts

\usepackage{sty/style}

\definecolor{crabred}{HTML}{BE1831}

\def\BibTeX{{\rm B\kern-.05em{\sc i\kern-.025em b}\kern-.08em
    T\kern-.1667em\lower.7ex\hbox{E}\kern-.125emX}}

\begin{document}

\title{A Physically Consistent Assessment of Nearfield Beamfocusing into Occluded  Regions} 

\author{\IEEEauthorblockN{Georg Schwan, Alexander Stutz-Tirri, and Christoph Studer}\thanks{GS, AST, and CS are with the Department of Information Technology and Electrical Engineering at ETH Zurich, Switzerland; email: gschwan@ethz.ch, alstutz@ethz.ch, and studer@ethz.ch}\thanks{This work was funded in part by armasuisse and by the Swiss State Secretariat for Education, Research, and Innovation (SERI) under the SwissChips initiative. This work has also received support by the Swiss National Science Foundation (SNSF) through the Project ``Ubiquitous Large InTelligent ArRAys (ULTRA)'' under Grant 219710 and the HORIZON-JU-SNS-2024-STREAM-C ``X-TREME 6'' Project under Grant 101192681.}
}

\maketitle


\begin{abstract}
Recent work has explored elaborate beamfocusing techniques for the radiative nearfield, with some studies suggesting that certain beamshapes, such as Airy beams, can enable efficient electromagnetic (EM) wave transmission behind obstacles. In this letter, we ask whether the added complexity of such techniques is justified. We distinguish between partially and fully occluded regions. In the partially occluded region, where Airy beams are commonly employed, we show that a simple line-of-sight (LoS) strategy, which activates only antennas having an unobstructed view of the receiver, is near-optimal and outperforms Airy beams at substantially lower complexity. In the fully occluded region, we argue that accurate beamfocusing requires a physically consistent EM wave propagation model that captures propagation effects such as diffraction. Once such a model is available, however, the optimal beamfocusing strategy has a closed-form solution and can be computed directly. These results suggest that elaborate  techniques, such as Airy beams, offer little benefit over simpler alternatives for beamfocusing into occluded regions. 

\end{abstract}

\section{Introduction}\label{sec:introduction}
Beamfocusing in the radiative nearfield region has received considerable attention in recent years, as it enables transmission to (and reception from) specific \emph{points} in space using a single antenna array~\cite{special_issue_on_near_field_singal_processing}.
Several works investigate more elaborate beamfocusing techniques.
A prominent example is the Airy beam~\cite{airy_beam_observation}, an
electromagnetic~(EM) beam whose main lobe follows a curved trajectory in
free space. This property has been proposed as a means to focus energy
around corners into \emph{partially occluded}
regions~\cite{zhao2026airybeamengineeringnearfield}.
Some recent studies go even further, suggesting that such specialized beam shapes may allow signals to be transmitted meaningfully into \emph{fully occluded} regions\mbox{\cite{liu2026bendingbeamthzwireless, qin2026airybeamformingradiativenearfield}}.

\subsection{Contributions}
In this letter, we ask whether Airy beams are a practicable technique to focus energy into partially or even fully occluded regions. Concretely, we investigate whether computationally less complex strategies can match or even exceed their performance. 
We investigate this question using the physically consistent EM nearfield model proposed in~\cite{schwan2026physicallyconsistentevaluationcommonly}.
Specifically, we consider a uniform linear array of $64$ patch antennas with an obstacle in its nearfield region, as depicted in~\fref{fig:illustration}, and we attempt to deliver energy into the partially and fully occluded regions behind this obstacle.
We compare four beamfocusing techniques: (i)~a na\"ive strategy, (ii) the Airy-based strategy of~\cite{chen_a_physics_informed_airy_beam_learning_framework}, (iii) an optimal strategy, and (iv) a line-of-sight (LoS) strategy; the latter two techniques are proposed in this work.

\def\ground (#1,#2) {
    \draw[shift={(#1,#2)}, line width=.6] (-0.16,0) -- (0.16,0);
    \draw[shift={(#1,#2)}, line width=.6] (-0.12,-0.05) -- (0.12,-0.05);
    \draw[shift={(#1,#2)}, line width=.6] (-0.08,-0.1) -- (0.08,-0.1);
}

\def\impedanceName (#1,#2,#3) {
    \draw  [shift={(#1,#2)}, line width=.7pt, anchor=center, fill=white] (-.3,-.12) rectangle (.3,.12);
    \node[shift={(#1,#2)}, anchor=south] at (0,0.12) {#3};
}

\def\impedance (#1,#2) {
    \draw  [shift={(#1,#2)}, line width=.7pt, anchor=center, fill=white] (-.3,-.12) rectangle (.3,.12);
}

\def\voltage (#1,#2) {
    \draw[shift={(#1,#2)}, line width=.7pt] (0,0) ellipse (0.28 and 0.28);
    \draw[shift={(#1,#2)}, line width=.7pt] (0,-.28) -- (0,.28);
}

\def\cont (#1,#2,#3) {
    \draw  [shift={(#1,#2)}, line width=.5pt, fill=grey, color=grey] (-.2,.14)  ellipse (0.02 and 0.02);
    \draw  [shift={(#1,#2)}, line width=.5pt, fill=grey, color=grey] (-.2,0) ellipse (0.02 and 0.02);
    \draw  [shift={(#1,#2)}, line width=.5pt, fill=grey, color=grey] (-.2,-0.14) ellipse (0.02 and 0.02);
    \node[shift={(#1,#2)}, anchor=west] at (-.2,0) {\color{grey} \footnotesize #3};
}

\def\antenna (#1,#2) {
    \draw[shift={(#1,#2)}, line width=.6] (0,-0.2) -- (0,0.4) -- (0.15,0.4) -- (0,0.1) -- (-0.15,0.4) -- (0,0.4);
}

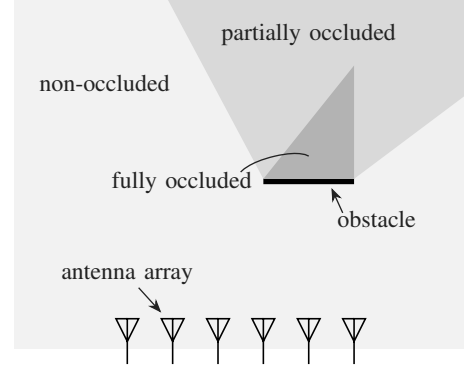
\begin{figure}
    \centering
    \small
    \begin{tikzpicture}
    
        \def\aperature{3};
        
        \def\heightclip{0.8 * \aperature};

        \def\leftclip{0.5 * \aperature};

        \fill[fill=gray!10] (-\aperature,0) -- (\aperature, 0) -- (\aperature,0.75*\aperature +\heightclip) -- (-\aperature,0.75*\aperature +\heightclip) -- cycle;

        \draw[line width=4.0] (0.1*\aperature,0.75*\aperature) -- (0.5*\aperature,0.75*\aperature);

        \fill[fill=gray!30] (0.1*\aperature,0.75*\aperature) -- (0.1*\aperature - \heightclip / 1.875,0.75*\aperature +\heightclip) -- (0.5*\aperature + \leftclip, 0.75*\aperature +\heightclip) -- (0.5*\aperature + \leftclip,0.75*\aperature + \leftclip * 0.75) -- (0.5*\aperature,0.75*\aperature) -- cycle;
        
        \fill[fill=gray!60] (0.5*\aperature,0.75*\aperature) -- (0.5*\aperature, 1.25*\aperature) -- (0.1*\aperature,0.75*\aperature) -- cycle;

        \antenna (0.5*\aperature, 0)
        \antenna (0.3*\aperature, 0)
        \antenna (0.1*\aperature, 0)
        \antenna (-0.1*\aperature, 0)
        \antenna (-0.3*\aperature, 0)
        \antenna (-0.5*\aperature, 0)

        \node[anchor=south] at (-0.6*\aperature, 1.1*\aperature) {\color{black_light}non-occluded};
        
        \node[anchor=south] at (0.3*\aperature, 1.3*\aperature) {\color{black_light}partially occluded};
        
        \node[anchor=south] at (-0.26*\aperature, 0.65*\aperature) {\color{black_light}fully occluded};
        
        \draw [color=black_light, line width=.5pt] plot[smooth, tension=1.4] coordinates {(0.00*\aperature, 0.78*\aperature)  (0.15*\aperature, 0.85*\aperature) (0.3*\aperature, 0.85*\aperature)};
        
        \node[anchor=south] at (0.6*\aperature, 0.5*\aperature) {\color{black_light}obstacle};
    
        \draw [-{Stealth}, color=black_light, line width=.5pt] plot[smooth, tension=1.4] coordinates {(0.45*\aperature, 0.6*\aperature) (0.4*\aperature, 0.71*\aperature)};
        
        \node[anchor=south] at (-0.5*\aperature, 0.25*\aperature) {\color{black_light}antenna array};
        \draw [-{Stealth}, color=black_light, line width=.5pt] plot[smooth, tension=1.4] coordinates {(-0.45*\aperature, 0.27*\aperature) (-0.35*\aperature, 0.16*\aperature)};
    
    \end{tikzpicture}
\caption{Illustration of the antenna array with an obstacle in its nearfield region. In the non-occluded nearfield region (light gray), every antenna has a LoS path to any point in the region; in the partially occluded nearfield region (medium gray), at least one but not all antennas have a LoS path; and in the fully occluded nearfield region (dark gray), no antenna has a LoS path.}

\label{fig:illustration}
\end{figure}

\blankfootnote{
{\em Notation:} We denote general vectors and matrices using lowercase (e.g.,~$\vect{a}$) and uppercase (e.g.,~$\mat{A}$) boldface, respectively.
For phasors (cf.~\cite[Def.~1]{stutz_schwan_studer_efficient_and_physically_consistent_modeling_of_reconfigurable_electromagnetic_structures}) and phasor vectors, we use pink sans-serif (e.g.,~$\phs{a}$) and boldface (e.g.,~$\phv{a}$).
We indicate the transpose and conjugate transpose by the superscripts~$^\T$ and~$^\He$, respectively.
We use $\diag(\vect{a})$ for the diagonal matrix with the elements of the vector~$\vect{a}$ on the main diagonal, $\|\cdot\|_2$ for the Euclidean norm, and blackboard bold font for operators (e.g.,~$\mathbb{S}$).
We select the $i$th element of the vector~$\vect{a}$ via $[\vect{a}]_i$ and denote the element-wise complex conjugate as~$\overline{\vect{a}}$.
We furthermore denote the element-wise real part of a complex-valued scalar,
vector, or matrix~$c$ by~$\Re\{c\}$. 
At frequency $f$, we define the free-space wavenumber as $k\triangleq 2\pi f\sqrt{\mu_\up{0}\varepsilon_\up{0}}$, free-space impedance as $Z_\up{0}\triangleq\sqrt{\mu_\up{0}/\varepsilon_\up{0}}$, and free-space wavelength as $\lambda\triangleq (\sqrt{\varepsilon_\up{0} \mu_\up{0}}f)^{-1}$, using the permeability $\mu_\up{0}$ and permittivity~$\varepsilon_\up{0}$ of free space.
We adopt the physicist's spherical coordinates~\cite{ISO_quantities_and_units_2_mathematics}~$(r,\,\theta,\,\varphi)$ representing radial distance, polar angle, and azimuthal angle.
}

\section{Physically Consistent Nearfield Model}\label{sec:framework}

We utilize the physically consistent nearfield model proposed in~\cite[Sec.~II]{schwan2026physicallyconsistentevaluationcommonly} to predict the EM field at a discrete set of prespecified coordinates. The model parameters are obtained from a single full-wave simulation.
We now restate the concepts required in this work and refer the reader to~\cite{schwan2026physicallyconsistentevaluationcommonly} for more details.

\subsection{Sampled Nearfield Model}
Let~$\phv{E}^\nearrow(r,\theta,\varphi)$ and $\phv{H}^\nearrow(r,\theta,\varphi)$ denote the phasor vectors of the outgoing electric and magnetic fields, respectively, i.e., the field components radiated directly by the antennas.\footnote{The model in~\cite{schwan2026physicallyconsistentevaluationcommonly} also accounts for incoming waves, which is why it distinguishes between the outgoing and the total EM field. Since, in this work, we consider outgoing waves only, the two coincide here.} 
In order to characterize the field in the region $\mathcal{V}^\up{c}$ exterior to the antennas (which includes the nearfield region), we define the \emph{outgoing EM field} vector as
\begin{align}
    \phv{a}_\up{N}\colon\mathcal{V}^\up{c}\rightarrow\mathbb{C}^{6},
    \quad
    (r,\theta,\varphi)\mapsto
    \begin{bmatrix}
        \tfrac{1}{\sqrt{Z_\up{0}}}\,\phv{E}^\nearrow(r,\theta,\varphi)
        \\[4pt]
        \sqrt{Z_\up{0}}\,\phv{H}^\nearrow(r,\theta,\varphi)
    \end{bmatrix}\!.
    \label{eq:definition_far_field_power_waves_a}
\end{align}

We model the internal behavior of the transmitter using three subsystems: the RF~frontend, the tuning network, and the radiating structure~\cite{schwan2026physicallyconsistentevaluationcommonly}. For brevity, we describe only the RF frontend, which comprises $N\in\mathbb{Z}_{>0}$ power amplifiers (PAs), each modeled by its Th\'evenin-equivalent circuit with the following voltages and impedances:
\begin{align}
        \phv{v}_\up{Tx}
        &\triangleq
        [\phs{v}_{\up{Tx},1}\;
        \cdots\;
        \phs{v}_{\up{Tx},N}]^\T
    \\
        \mat{Z}_\up{Tx}
        &\triangleq
        \diag\big([\mathrm{Z}_{\up{Tx},1}\;
        \cdots\;
        \mathrm{Z}_{\up{Tx},N}]\big).
\end{align}
Assuming that the transmitter can be modeled as a linear time-invariant (LTI) system, the relation between the PA voltage (phasor) vector~$\phv{v}_\up{Tx}$ and the outgoing EM field (phasor) vector~$\phv{a}_\up{N}$ is linear and given by 
\begin{align}
    \phv{a}_\up{N}
    &=
    \mathbb{G}_{\phv{v}_\up{Tx}}^{\phv{a}_\up{N}}\,\phv{v}_\up{Tx},
    \label{eq:input_output}
\end{align}
with the gain operator $\mathbb{G}_{\phv{v}_\up{Tx}}^{\phv{a}_\up{N}}$
defined in~\cite[Eq.~10]{schwan2026physicallyconsistentevaluationcommonly}.

\begin{rem}\label{rem:obstacle}
In our model, we account for obstacles by treating them as part of the antenna system. Their scattering response is thereby absorbed into the gain operator $\mathbb{G}_{\phv{v}_\up{Tx}}^{\phv{a}_\up{N}}$. 
\end{rem}

\begin{rem}
Because the outgoing EM field vector~$\phv{a}_\up{N}$ is defined over a continuum
of coordinates, the codomain of the gain
operator~$\mathbb{G}_{\phv{v}_\up{Tx}}^{\phv{a}_\up{N}}$ is infinite-dimensional,
and the operator therefore does not admit a finite-dimensional representation in
general.
One key idea in~\cite{schwan2026physicallyconsistentevaluationcommonly} is to
sample the operator~$\mathbb{G}_{\phv{v}_\up{Tx}}^{\phv{a}_\up{N}}$ at finitely
many coordinates in space.
Specifically, we denote the set of sample points by~$\tilde{\mathcal{V}}^\up{c}\subset\mathcal{V}^\up{c}$,
and for each~$(r,\theta,\varphi)\in\tilde{\mathcal{V}}^\up{c}$, we implicitly
define the matrix~$\mat{G}_{\phv{v}_\up{Tx}}^{\phv{a}_\up{N}}(r,\theta,\varphi)\in\mathbb{C}^{6\times N}$ as
\begin{align}
    \phv{a}_\up{N}(r,\theta,\varphi)=\mat{G}_{\phv{v}_\up{Tx}}^{\phv{a}_\up{N}}(r,\theta,\varphi)\phv{v}_\up{Tx}.
\end{align}
Note that these finitely many matrices can be extracted from a single full-wave
EM simulation.
\end{rem}

\subsection{Power-Normalized Energy Density}
In order to carry out a receive-antenna-independent analysis, we consider the
energy density of the transmitted EM field.\footnote{Note that the energy density can be a meaningful proxy for the power captured by an
electrically small antenna.} 
The \emph{outgoing energy density} of the EM field at $(r,\theta,\varphi)$ is given by~\cite[Eq.~9.71]{griffiths_introduction_to_electrodynamics}
\begin{align}
    u(\phv{v}_\up{Tx};r,\theta,\varphi)
    &\triangleq
    \frac{1}{2}\big(\varepsilon_\up{0}\|\phv{E}^\nearrow\|_2^2 + \mu_\up{0}\|\phv{H}^\nearrow\|_2^2\big)
    \label{eq:energy_density_fields} \\
    &=
    \frac{\sqrt{\mu_\up{0}\varepsilon_\up{0}}}{2}\,\|\phv{a}_\up{N}(r,\theta,\varphi)\|_{2}^2.
    \label{eq:energy_density}
\end{align}

In order to prevent any scheme from maximizing the outgoing energy density by
simply increasing the PAs' output power, we furthermore normalize the outgoing
energy density~$u$ by a power metric. Specifically, we utilize the \emph{PA available power} introduced in~\cite[Def.~6]{stutz_schwan_studer_efficient_and_physically_consistent_modeling_of_reconfigurable_electromagnetic_structures}, which, for a fixed PA voltage vector $\phv{v}_\up{Tx}\in\mathbb{C}^N$, is given by\footnote{Note that other power metrics could also be meaningfully used for normalization (cf.~\cite[Sec.~II-D]{ferencikova_schwan_joint_beamforming_and_matching}). For
example, one could use the PAs' actual output power instead of the available power~$P_\up{A}$.} 
\begin{align}
    \label{eq:defi_PA}
    P_\up{A}
    =
    \frac{1}{4}\,\phv{v}_\up{Tx}^\He\,
    \Re\{\mat{Z}_\up{Tx}\}^{-1}\,
    \phv{v}_\up{Tx}.
\end{align}

With these definitions, we now propose the \emph{power-normalized energy density}, which, for a given PA voltage vector $\phv{v}_\up{Tx}\in\mathbb{C}^N$ and coordinate~$(r,\theta,\varphi)$, is defined as
\begin{align}
    u_\up{A}(\phv{v}_\up{Tx};r,\theta,\varphi)
    \triangleq
    \frac{u(\phv{v}_\up{Tx};r,\theta,\varphi)}{P_\up{A}}.
    \label{eq:normalized_power_energy_density}
\end{align}

\section{Beamfocusing}\label{sec:beamfocus}
We now detail the four beamfocusing strategies, which we use for our investigations in \fref{sec:results}. 

\subsection{Na\"ive Strategy}
The na\"ive beamfocusing strategy ignores the presence of obstacles and uses the spherical-wave-based model from~\cite[Sec.~II-B2]{yuanwei_near_field_communication_a_tutorial_review} for beamfocusing.\footnote{As shown in~\cite{schwan2026physicallyconsistentevaluationcommonly}, the spherical-wave-based model agrees well with the physically consistent model for a uniform linear array with $\lambda/2$ inter-antenna spacing in the absence of obstacles.}
The na\"ive beamfocusing vector is defined as
\begin{align}
    \phv{v}^\up{na\"ive}_\up{Tx}(r^\star, \theta^\star,\varphi^\star)
    \triangleq\, 
    \overline{\vect{c}}(r^\star, \theta^\star,\varphi^\star),
\end{align}
where $\vect{c}(r, \theta,\varphi)\in\mathbb{C}^{N}$ is given by~\cite[Eq.~36]{yuanwei_near_field_communication_a_tutorial_review}
\begin{align}\label{eq:array_factor}
    \vect{c}(r, \theta,\varphi)
    \triangleq
    \begin{bmatrix}
    \dfrac{e^{-jkd_1(r, \theta,\varphi)}}{d_1(r, \theta,\varphi)} & \cdots & \dfrac{e^{-jkd_N(r, \theta,\varphi)}}{d_N(r, \theta,\varphi)}
    \end{bmatrix}^\T\!,
\end{align}
with $d_i(r, \theta,\varphi)$ denoting the distance between the coordinate~$(r, \theta,\varphi)$ and the $i$th antenna element.
\begin{rem}
The na\"ive beamfocusing vector is straightforward to compute, as it requires only the geometry of the antenna array and the focus coordinate.
\end{rem}

\subsection{Airy Strategy}
An Airy beam is a wave whose main intensity lobe propagates in free space along a curved, parabolic trajectory~\cite{airy_beam_observation}. Airy beams can be parametrized by the triple $(B, F, \alpha) \in \mathbb{R}^3$ of curvature, focal length, and steering angle. The Airy beam can be synthesized by imposing the following phase profile at the transmitter~\cite[Eq.~2]{chen_a_physics_informed_airy_beam_learning_framework}:
\begin{align}
& \phi(x;B,F,\alpha) = \notag \\
& \qquad \tfrac{1}{3}(2\pi B )^3x^3 - k\sqrt{(F\sin\alpha - x)^2 + (F\cos\alpha)^2},
\end{align}
where $x$ denotes the position along the array. The set $\mathcal{A}$ of possible phase configurations for a linear antenna array is
\begin{align}
   \bigl\{ \phv{v} \in \mathbb{C}^N \mid
    \exists\, (B,F,\alpha) \text{ s.t. } [\phv{v}]_i = e^{-j\phi(x_i;B,F,\alpha)}\ \forall i \bigr\},
\end{align}
where $x_i$ is the position of the $i$th antenna element. The optimal Airy beamfocusing vector is then a solution to
\begin{align}
    \phv{v}^\up{Airy}_\up{Tx}(r^\star, \theta^\star,\varphi^\star)
    \in
    \argmax_{\phv{v}_\up{Tx} \in \mathcal{A}}\; u_\up{A}(\phv{v}_\up{Tx};r^\star, \theta^\star,\varphi^\star).
\end{align}

\begin{rem}
This optimization problem is difficult to solve in general.
We therefore approximate its solution by a fine grid search over the parameters $(B,F,\alpha)$, which we subsequently refine using gradient ascent.
Note that this strategy requires knowledge of the gain matrix~$\mat{G}_{\phv{v}_\up{Tx}}^{\phv{a}_\up{N}}(r^\star,\theta^\star,\varphi^\star)$.
Note that the closed-form parameter selection used in~\cite{zhao2026airybeamengineeringnearfield} did not yield competitive solutions; we therefore rely on the numerical procedure above to give the Airy strategy a fair comparison.
\end{rem}

\subsection{Optimal Strategy}
The optimal beamfocusing vector, which maximizes the power-normalized energy density at the focus coordinate~$(r^\star, \theta^\star,\varphi^\star)$, is a solution to the optimization problem
\begin{align}
    \label{eq:optimal_beam_problem}
    \phv{v}^\up{opt}_\up{Tx}(r^\star, \theta^\star,\varphi^\star)
    \in
    \argmax_{\phv{v}_\up{Tx} \in \mathbb{C}^N\setminus\{\vect{0}\}}\; u_\up{A}(\phv{v}_\up{Tx}; r^\star, \theta^\star,\varphi^\star).
\end{align}
This is a generalized Rayleigh-quotient maximization, whose closed-form solution is given by~\cite[Sec.~III]{ferencikova_schwan_joint_beamforming_and_matching}
\begin{align}
    \label{eq:solution_optimal_beam}
    \phv{v}^\up{opt}_\up{Tx}(r^\star, \theta^\star,\varphi^\star)
    \triangleq
    \Re\{\mat{Z}_\up{Tx}\}^{\frac{1}{2}}\,
    \vect{q}(r^\star, \theta^\star,\varphi^\star),
\end{align}
where $\vect{q}(r^\star, \theta^\star,\varphi^\star)$ is the dominant eigenvector of the matrix \mbox{$\Re\{\mat{Z}_\up{Tx}\}^{\frac{1}{2}}\big(\mat{G}_{\phv{v}_\up{Tx}}^{\phv{a}_\up{N}}(r^\star,\theta^\star,\varphi^\star)\big)^\He\mat{G}_{\phv{v}_\up{Tx}}^{\phv{a}_\up{N}}(r^\star,\theta^\star,\varphi^\star)\Re\{\mat{Z}_\up{Tx}\}^{\frac{1}{2}}$}.

\begin{rem}\label{rem:airy}
    The optimal beamfocusing strategy achieves the maximum possible
    power-normalized energy density at the focus coordinate~$(r^\star,\theta^\star,\varphi^\star)$; \emph{no} other beamfocusing strategy can outperform it.
    This strategy, however, requires knowledge of the gain
    matrix~$\mat{G}_{\phv{v}_\up{Tx}}^{\phv{a}_\up{N}}(r^\star,\theta^\star,\varphi^\star)$---as
    does the Airy strategy---which can be extracted from a full-wave EM simulation.
\end{rem}

\subsection{Line-of-Sight Strategy}

We now propose what we call the LoS strategy. The idea is simple: We follow the na\"ive strategy but use only those antennas that have a LoS path to the focus coordinate, disabling the remaining antennas. The LoS beamfocusing vector is given by
\begin{align}
\big[\phv{v}^\up{LoS}_\up{Tx}(r^\star, \theta^\star,\varphi^\star)\big]_i
\triangleq
\begin{cases}
\big[\overline{\vect{c}}(r^\star, \theta^\star,\varphi^\star)\big]_i \!\! &\!\! \text{if antenna } i \text{ has LoS}, \notag \\
0 & \! \text{otherwise.}
\end{cases}  \\[-0.55cm]
\end{align}

\begin{rem}
The LoS beamfocusing vector is straightforward to compute, as it requires only knowledge of the antenna geometry, the obstacle position, and the focus coordinate.
\end{rem}

\section{Results}\label{sec:results}
\begin{figure}
    \centering
    \includegraphics[width=0.99\columnwidth]{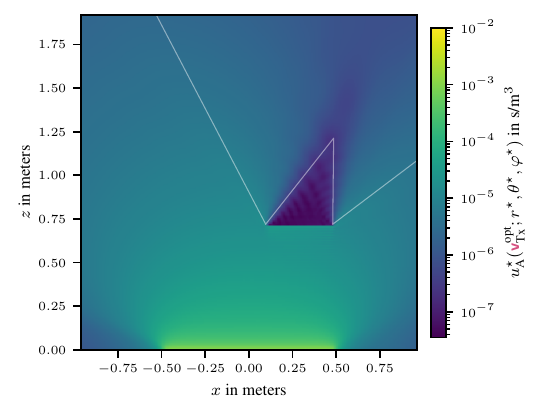}
    \vspace{-0.6cm}
\caption{Power-normalized energy density $u^\star_\up{A}(\phv{v}^\up{opt}_\up{Tx};\, r^\star, \theta^\star,\varphi^\star)$ achieved by the \emph{optimal} beamfocusing vector at each coordinate, which upper-bounds the energy density attainable at that coordinate. The white lines mark the boundaries of the different occluded regions (cf.~\fref{fig:illustration}). Behind the obstacle, the energy density drops sharply, indicating that any beamfocusing strategy can deliver only a small amount of energy into the fully occluded region.}
\label{fig:plot_optimal}
\end{figure}

\begin{figure*}[tbh]
    \centering
    \includegraphics[width=0.99\textwidth]{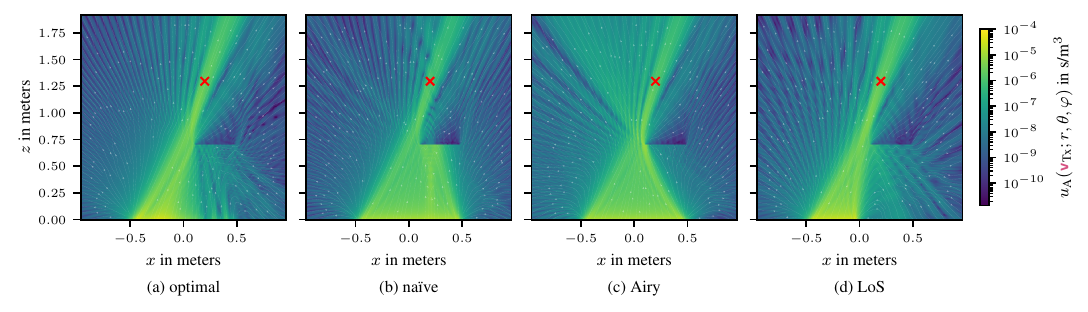}
    \vspace{-0.3cm}
    \caption{Power-normalized energy density $u_\up{A}(\phv{v}_\up{Tx};r,\theta,\varphi)$ across space for the four beamfocusing strategies described in Section~\ref{sec:beamfocus}. The focus coordinate (marked with a red $\textcolor{red}{\times}$) lies in the partially occluded region at $(x,z) = (\SI{0.2}{\meter},\,\SI{1.3}{\meter})$. The white lines indicate the direction of the time-averaged Poynting vector~\cite[Eq.~8.10]{griffiths_introduction_to_electrodynamics}, i.e., the direction of power flow. The shape of each beam is clearly visible.}
\label{fig:plot_partially_occluded}
\end{figure*}

\begin{figure*}[tbh]
    \centering
    \includegraphics[width=0.99\textwidth]{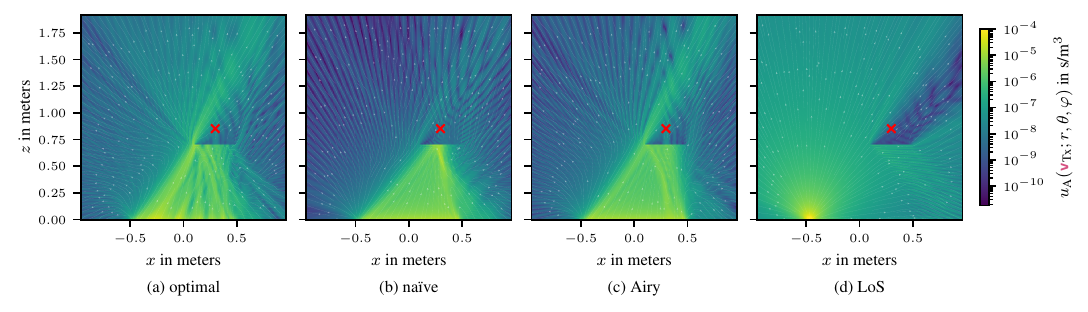}
    \vspace{-0.3cm}
    \caption{Power-normalized energy density $u_\up{A}(\phv{v}_\up{Tx};r,\theta,\varphi)$ across space for the four beamfocusing strategies described in Section~\ref{sec:beamfocus}. The focus coordinate (marked with a red $\textcolor{red}{\times}$) lies in the fully occluded region at $(x,z) = (\SI{0.3}{\meter},\,\SI{0.8}{\meter})$. The white lines indicate the direction of the time-averaged Poynting vector~\cite[Eq.~8.10]{griffiths_introduction_to_electrodynamics}, i.e., the direction of power flow. None of the strategies produces a visible energy peak at the focus coordinate.}
\label{fig:plot_fully_occluded}
\end{figure*}

\begin{figure*}
    \centering
    \includegraphics[width=0.99\textwidth]{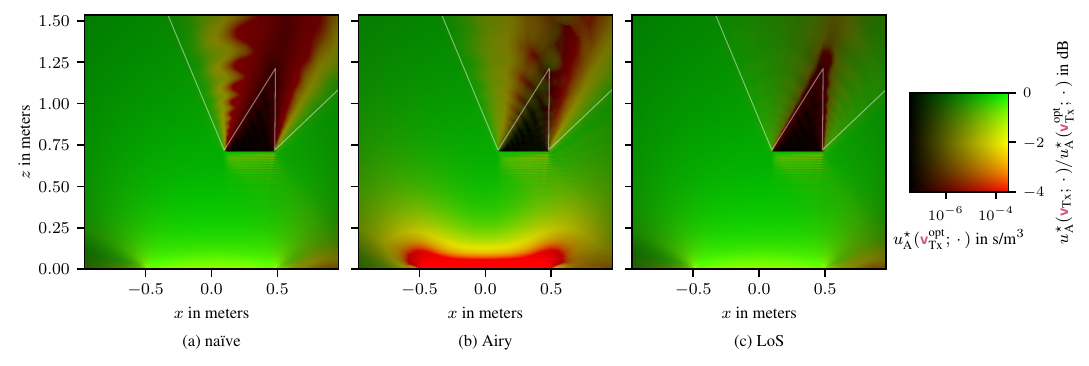}
    \vspace{-0.3cm}
    \caption{Performance comparison of the na\"ive, the Airy, and the LoS strategies against the optimal strategy at each coordinate. The hue encodes the ratio~$u^\star_\up{A}(\phv{v}_\up{Tx};\, r^\star, \theta^\star,\varphi^\star)/u^\star_\up{A}(\phv{v}^\up{opt}_\up{Tx};\, r^\star, \theta^\star,\varphi^\star)$, where green indicates near-optimal performance and red poor performance. The brightness encodes the value of~$u^\star_\up{A}(\phv{v}^\up{opt}_\up{Tx};\, r^\star, \theta^\star,\varphi^\star)$, i.e., how much energy the best possible strategy can deliver. The white lines mark the boundaries of the different occluded regions (cf.~\fref{fig:illustration}). The simple LoS strategy is already sufficient to focus the energy density effectively in the partially occluded region.}
    \label{fig:plot_compare}
\end{figure*}

We now compare how much energy the four beamfocusing strategies deliver across different focus coordinates.

\subsection{Setup}
We consider a uniform linear array of $N=64$ patch antennas operating at \SI{10}{\giga\hertz} with an inter-antenna spacing of $\lambda/2$. The array is centered at the origin in the $z=\SI{0}{\meter}$ plane. The obstacle is an infinitely thin, perfectly conducting metal plate spanning from $(x,z) = (\SI{0.096}{\meter},\,\SI{0.72}{\meter})$ to $(\SI{0.48}{\meter},\,\SI{0.72}{\meter})$.\footnote{Along the $y$-axis, the plate is sufficiently wide for edge effects in that direction to be neglected.} The PA equivalent impedances are $\mat{Z}_\up{Tx} = \SI{50}{\ohm}\cdot\mat{I}_N$. We obtained the gain matrix $\mat{G}_{\phv{v}_\up{Tx}}^{\phv{a}_\up{N}}$ with Ansys HFSS~\cite{ansys_hffs} at the single frequency of \SI{10}{\giga\hertz}.\footnote{Repeating this at multiple frequencies would enable a wideband analysis; we leave such an analysis for future work.}

\subsection{Optimal Performance}
We start by evaluating the energy density across all focus coordinates for the optimal beam. To simplify the notation, we introduce the shorthand\footnote{We slightly abuse notation: in this context, $\phv{v}_\up{Tx}$ does not denote a fixed vector, but rather a function $\phv{v}_\up{Tx}\colon\mathcal{V}^\up{c}\rightarrow\mathbb{C}^{N}$ that assigns a voltage vector to every coordinate; the function acts as a placeholder for any one of $\phv{v}^\up{opt}_\up{Tx}$, $\phv{v}^\up{na\"{\i}ve}_\up{Tx}$, $\phv{v}^\up{Airy}_\up{Tx}$, and $\phv{v}^\up{LoS}_\up{Tx}$.}
\begin{align}
    u^\star_\up{A}(\phv{v}_\up{Tx};\, r^\star, \theta^\star,\varphi^\star)
    \triangleq
    u_\up{A}\big(\phv{v}_\up{Tx}(r^\star, \theta^\star,\varphi^\star);\, r^\star, \theta^\star,\varphi^\star\big),
\end{align}
which denotes the power-normalized energy density at coordinate $(r^\star,\theta^\star,\varphi^\star)$ when the beam is focused on that coordinate.

\fref{fig:plot_optimal} shows $u^\star_\up{A}(\phv{v}^\up{opt}_\up{Tx};\, r^\star, \theta^\star,\varphi^\star)$ for the optimal beamfocusing vector at each coordinate.
Behind the obstacle, in the fully occluded region, the energy density drops sharply, by more than \SI{20}{\decibel}.
Since the optimal strategy upper-bounds the energy density achievable by any other strategy, this demonstrates that \emph{any} beamfocusing strategy can deliver only a small amount of energy into the fully occluded region.
The reason is that, in the considered scenario, diffraction is the only mechanism by which EM waves reach the fully occluded region.\footnote{In general, reflection, refraction, scattering, penetration, and guided propagation also enable waves to reach occluded regions.\label{fn:refraction}}
Consequently, the energy reaching the fully occluded region decreases even further at higher carrier frequencies, where diffraction is less pronounced~\cite[Ch.~10]{hecht_optics}.

\subsection{Beam Shapes}
To build intuition for how the beams behave, we fix a single focus coordinate and examine the resulting power-normalized energy density across space. We consider two such focus coordinates: (i) a coordinate in a partially occluded region, located at $(x,z) = (\SI{0.2}{\meter},\,\SI{1.3}{\meter})$, and (ii) a coordinate in a fully occluded region, located at $(x,z) = (\SI{0.3}{\meter},\,\SI{0.8}{\meter})$.

\fref{fig:plot_partially_occluded} shows the resulting power-normalized energy density for each beamfocusing strategy in the partially occluded case. We can see that for the na\"ive strategy, a substantial fraction of the energy is reflected by the metal plate. For the Airy strategy, the characteristic curvature of the main lobe is clearly visible, especially between the array and the obstacle. For the optimal strategy, most of the antennas with a direct LoS path to the focus coordinate are active. Note that the LoS strategy is visually the closest to the optimal beam. 

\fref{fig:plot_fully_occluded} shows the power-normalized energy density for each strategy in the fully occluded case.\footnote{Since no antenna has a LoS path to the focus coordinate, the LoS beamfocusing vector is all-zero, for which $u_\up{A}$ is undefined; we therefore activate only the leftmost antenna.}
We observe that none of the strategies produces a visible energy peak at the focus coordinate. The reason is again that, apart from diffraction, \emph{no} EM wave reaches the fully occluded region in this scenario.

\subsection{Performance Comparison}
We now compare the beamfocusing performance quantitatively. Rather than just the two focus coordinates considered above, we now again evaluate and compare the energy density at all focus coordinates in the considered region. 

\fref{fig:plot_compare} uses a two-dimensional color encoding. The hue encodes the ratio $u^\star_\up{A}(\phv{v}_\up{Tx};\, r^\star, \theta^\star,\varphi^\star)/u^\star_\up{A}(\phv{v}^\up{opt}_\up{Tx};\, r^\star, \theta^\star,\varphi^\star)$, i.e., the
energy density of a given strategy relative to that of the optimal beam; the brightness encodes the denominator $u^\star_\up{A}(\phv{v}^\up{opt}_\up{Tx};\, r^\star, \theta^\star,\varphi^\star)$, i.e., how much energy can be delivered to that
coordinate at all. 
In words: green color implies similar performance to the optimal beam; red color implies worse performance than the optimal beam; black implies that no strategy can deliver energy to that area. 
This lets us compare each strategy against the optimal beam while simultaneously
showing how much energy can propagate into a given region. 

We see that, in the non-occluded region, all strategies approximately match the optimal strategy.\footnote{An exception is the Airy beam in the immediate vicinity of the array, a region of little practical relevance for Airy beams.} In the partially occluded region, the na\"ive and Airy beams degrade with the degree of occlusion, whereas the LoS beam remains near-optimal. This suggests that the proposed simple LoS beamfocusing strategy is already sufficient to focus energy effectively in partially occluded regions. 
In the fully occluded region, the Airy beam approaches the performance of the optimal beam, which indicates that the set of realizable Airy beams is rich enough here to come close to the optimum.
This fully occluded region, however, is precisely where our results imply that a physically consistent model is indispensable for parameter tuning (cf.~\fref{rem:airy}). However, once such a model is available, the optimal beam can be computed directly. Thus, the Airy beam offers no advantage: it requires the same model while being more computationally complex to tune.

\begin{rem}
We optimized the Airy beam parameter triple $(B, F, \alpha)$ using our physically consistent model. Therefore, the Airy strategy performance presented in this work represents a best-case scenario; we have not analyzed how much its performance would degrade if the parameters were tuned without a physically consistent model.
\end{rem}

\section{Conclusions}\label{sec:conclusions}
We have used a physically consistent EM model to investigate whether
elaborate beamfocusing techniques, such as those based on Airy beams,
justify their tuning complexity.
For the considered free-space scenario with a single obstacle, we have shown
that in the partially occluded region our proposed LoS beamfocusing strategy
already attains near-optimal performance, without resorting to intricate
beam shapes that are difficult to tune in practice.
In the fully occluded region, we have argued that a physically consistent
model is indispensable, since effects such as diffraction must be captured
correctly.
Once such a model is available, however, the optimal beam can be computed in
closed form. The Airy beam therefore offers no advantage: it requires the
same model while being considerably harder to tune.
In summary, Airy-beam-based beamfocusing is \emph{not} worth its complexity in
either region.
We emphasize that, throughout the letter, we assumed that channel reciprocity cannot be exploited. If, however, uplink pilots are available, reciprocal beamforming would be the better approach as it avoids parameter tuning altogether.

We conclude by noting that elaborate beamfocusing techniques, such as Airy beams, may nonetheless be attractive for other reasons, such as self-healing~\cite{Broky:08} or the fact that purely controlling the phase allows the PAs to operate at constant output power; we leave an investigation of such aspects for future work.
Further open questions include the analysis of scenarios with obstacles of different shapes and positions, as well as a comparison with other beam types, such as Gaussian, Bessel, and nonparaxial accelerating beams of the Mathieu and Weber type~\cite{PhysRevLett.109.193901}. We also leave to future work the impact of amplitude tapering, efficient methods for selecting beamfocusing parameters from a physically consistent model, and the extension to wideband operation.

We will make the model parameters and our code publicly available upon acceptance of this letter.

\balance
\bstctlcite{IEEEexample:BSTcontrol} 
\bibliographystyle{IEEEtran}
\bibliography{bib/publishers,bib/journals_proceedings_ect,bib/library}
\balance

\end{document}